\documentclass{aa}  

\usepackage{graphicx}
\usepackage{upgreek}
\usepackage{txfonts}
\usepackage{adjustbox}
\usepackage{color}
\usepackage{amsmath}
\usepackage{amssymb}
\usepackage{breqn}
\usepackage{gensymb}
\usepackage{placeins}
\usepackage{natbib}
\usepackage{subfig}
\usepackage{multirow}
\usepackage{booktabs}
\usepackage{colortbl}
\usepackage{csvsimple}
\usepackage[colorlinks=true,
linkcolor=blue,
citecolor=magenta]{hyperref}
\usepackage{txfonts}
\usepackage{comment}
\usepackage{gensymb}
\usepackage{pgfplots}
\newlength\figureheight
\newlength\figurewidth
\usepackage{soul}
\usepackage{orcidlink}

\usepackage{calligra}
\usepackage[T1]{fontenc}

\newcommand{\edit}[1]{#1}
\newcommand{\remove}[1]{}

\begin{document}

\title{Up around the bend: a multi-wavelength view of the quasar 3C\,345}

\titlerunning{A multi-wavelength view of 3C\,345}

\author{Jan R\"oder \orcidlink{0000-0002-2426-927X}\inst{\ref{corr}}\fnmsep
\thanks{Member of the International Max Planck Research School (IMPRS) for Astronomy and Astrophysics at the Universities of Bonn and Cologne} 
    \and
    Eduardo Ros \orcidlink{0000-0001-9503-4892} \inst{\ref{corr}} 
    \and
    Frank K. Schinzel \orcidlink{0000-0001-6672-128X}
    \inst{\ref{nrao}}
    \and
    Andrei P. Lobanov \orcidlink{0000-0003-1622-1484}
    \inst{\ref{corr}} 
    }

\authorrunning{J. R\"oder et al.}

\institute{
  Max-Planck-Institut f\"ur Radioastronomie, Auf dem H\"ugel 69, D-53121 Bonn, DE\\\email{jroeder.astro@gmail.com}\label{corr}
  \and
  National Radio Astronomy Observatory, P.O. Box O, Socorro, NM 87801, USA\label{nrao}
}

\date{Received 20 December, 2023; accepted 19 Feb, 2024}

\abstract
{The flat-spectrum radio quasar 3C\,345 has been showing gamma-ray \edit{activity} since the mid-2000s along with activity across the electromagnetic spectrum.
A gamma-ray burst in 2009 was successfully linked to relativistic outflow in 43\,GHz very long baseline interferometry (VLBI) observations and has since been analyzed also using single dish measurements. A multi-wavelength follow-up VLBI observation to the 2009 flare in conjunction with 43\,GHz catalogue data \edit{from the VLBA-BU-BLAZAR and BEAM-ME programs are} analyzed in this study in the context of the long-term evolution of the source.}
{We aim to probe the innermost few milli-arcseconds of the ultracompact 3C\,345 jet. In the process, we analyze the long-term kinematics of the inner jet and discuss the magnetic field morphology at different scales, as well as the origin of the gamma-ray emission.}
{New observations at 23, 43, and 86\,GHz took place on ten epochs between 2017 and 2019. We calibrate the 30 data sets using the \texttt{rPicard} pipeline, image them in \texttt{Difmap} and carry out polarization calibration using the \texttt{GPCAL} pipeline. We complement our VLBI data with single-dish lightcurves as well as ancillay VLBI maps at multiple frequencies. This data is then complemented by 43\,GHz observation carried out in the framework of the BEAM-ME and VLBA-BU-BLAZAR monitoring programs.}
{We find multiple distinct component paths in the inner jet, forming a helical geometry. The helix is anchored at a stationary feature some 0.16\,mas from the 43\,GHz VLBI core and has a timescale of about 8 years. The characteristic bends in the jet morphology are caused by variations in the component ejection angle. We confirm the result of previous studies that the gamma-ray emission is produced (or caused) by relativistic outflow and violent interactions within the jet. \remove{The results obtained in this paper will be used as input for relativistic magneto-hydrodynamic simulations in a follow-up study.}}
{}


\maketitle

\section{Introduction}

In recent years, very long baseline interferometry (VLBI) observations with 
a variety of telescope arrays have enabled us to study the formation and evolution of relativistic jets in great 
detail \citep[e.\,g.][]{Kim2020,Issaoun2022,Jorstad2022,Lee2013,Lee2014,Lee2016_I,Lee2016_II,Nair2019}. Among 
others, the flat-spectrum quasar 3C\,345 is one of the most extensively studied active galactic nuclei (AGN) \citep[e.\,g.,][]{Wittels1976_1,Wittels1976_2,Shapiro1979,Biretta1983,Biretta1986,Cohen1981,Cohen1983,Unwin1992,Zensus1995,Gabuzda1999,Klare2001,Lobanov1999,Lobanov2005,Ros2000,Schinzel2011}. 3C\,345 is monitored in the MOJAVE 15\,GHz 
\citep{Lister2021}, VLBA--BU--BLAZAR \& BEAM--ME 43\,GHz and SMA calibrator 230\,GHz programs 
\citep{Gurwell2007}, as well as by the FERMI/LAT gamma ray telescope \citep{Fermi2021,Abdollahi2023}.

At radio frequencies, 3C\,345 is known to show a characteristic core-jet structure with a notable persistent bend 
changing the jet direction from west to north-west at some 4\,mas from the core \citep{Schinzel2010}. The compact 
jet between the central engine and the northward bend consists of both stationary and moving shocks with 
transverse magnetic field orientation. 

The downstream shocks are frequently observed moving at apparent superluminal speeds \citep[e.\,g.,][]{Zensus1995,Lister2019,Lister2021,Weaver2022} 
and show a typical transverse magnetic field as expected from shock compression. At 15\,GHz, the magnetic field is longitudinal, 
indicative of a shear layer that is resolved out at higher frequencies (see Sec. \ref{sec:polarization}). The VLBI core keeps its 
transverse magnetic field in this frequency range. \cite{Lobanov1999} have postulated a $\sim$\,7.5 year activity cycle in 3C\,345, which for the most part has 
served somewhat accurate predictions of flares. 

The relativistic parsec-scale outflow has been identified as a source of the observed $\gamma$-ray emission in 3C\,345 \citep{Schinzel2011,Schinzel2012}, as well as other AGN \citep{MacDonald2017,Angioni2019,Angioni2020,KimDW2020,KimDW2022,Roesch2022}. Specifically, they showed a correlation between a newly ejected component passing through a stationary feature 
near the 43\,GHz core, and a spike in the $\gamma$-ray flux. This discovery prompted the proposal for the multi-frequency follow-up observations analyzed in this paper. 

Following \cite{Potzl2021}, we adopt a flat $\Lambda$CDM cosmology 
where $\Omega_M=0.3$, $\Omega_\Lambda=0.7$ and $H_0=70$\,km\,s$^{-1}$\,Mpc$^{-1}$. At a redshift $z=0.593$, this 
yields a linear scale of 6.65\,pc\,mas$^{-1}$ and a proper motion 
scale of $1{\rm \,mas\,yr^{-1}}$ corresponding to 34.5\,c. The paper 
is organized as follows: In section 2, the multi-frequency VLBA 
observations and data reduction are described. Section 3 is split 
into an analysis of the total intensity jet morphology, its 
kinematics, and its polarized structure. We discuss and summarize the results in section 4.

\section{Observations, calibration and imaging}
The VLBA campaigns BS260 and BS263 respectively took place on four 
observing days between June 27 and September 26, 2017 and six days 
between February 12 and August 28, 2018. The quasars 3C\,345 and 
3C\,279 were observed at central frequencies of 23.7, 43.2 and 
86.0\,GHz, along with the two calibrator sources J1310+3233 and J1407+2827 at the two lower frequencies.

The sources were observed with the whole VLBA with stations in Brewster (BR), Fort Davis (FD), Hancock (HN), Kitt Peak (KP), Los Alamos (LA), North Liberty (NL), Owens
Valley (OV), Pie Town (PT), Saint Croix (SC) and Mauna Kea (MK). Missing antennas for each epoch are reported in Table \ref{tab:obssummary}, such as the SC downtime between 2017 and 2018 due to hurricane damage. 
The data were recorded in two sub-bands (intermediate frequency channels) of 128\,MHz bandwith for both left-hand circular (LCP) and right-hand circular (RCP) polarization, and correlated at the VLBA correlator. An overview of the observations is given in Table \ref{tab:obssummary}.

The data were calibrated in total intensity using the \texttt{CASA}-based pipeline \texttt{rPicard} 
\citep{Janssen2019} and subsequently imaged in \texttt{Difmap} \citep{Shepherd1997}. Further, we used the 
\texttt{AIPS}-based pipeline \texttt{GPCAL} \citep{Park2021} and the \texttt{AIPS} tasks \texttt{IMAGR, COMB} and 
\texttt{PCNTR} \citep{Greisen2003} to calibrate and image the polarized emission. Due to the lack of a calibrator 
source, we are unable to reliably calibrate the EVPA. The 86\,GHz data are very noisy in some epochs, making 
polarization calibration infeasible and allowing only for Gaussian model fitting.

We complement this VLBA data set with 87 epochs at 43\,GHz taken from the VLBA-BU-BLAZAR and BEAM-ME 
programs, the first 46 of which are reported in \cite{Weaver2022}. \edit{Lastly, all gamma ray data are taken from the Fermi-LAT gamma-ray light curve repository \citep{Fermi2021,Abdollahi2023}\footnote{\href{https://fermi.gsfc.nasa.gov/ssc/data/access/lat/LightCurveRepository/}{https://fermi.gsfc.nasa.gov/ssc/data/access/lat/LightCurveRepository/}}}.

\section{Results and discussion}

\subsection{Total intensity jet structure}\label{sec:total_intensity}

\begin{figure}
    \centering
	\includegraphics[width=\columnwidth,trim={0 0 10 0},clip]{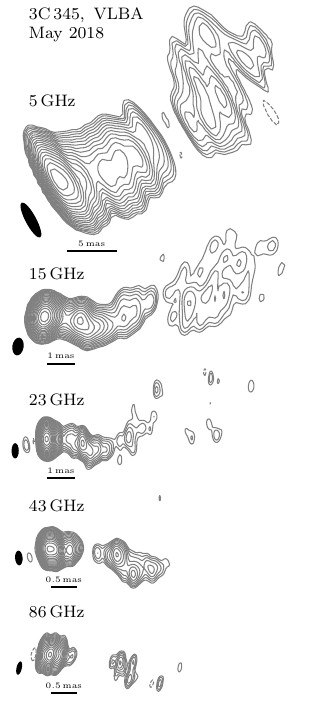}
    \vspace{-30pt}
	\caption{CLEAN images of the total intensity radiation in the jet of 3C\,345 at 5, 15, 23, 43, and 86 GHz, in uniform weighting. The restoring beam is shown in the bottom left corner of each image as black ellipse. Contour levels increase as $n\times2^{(i-2)/2}\times S_{\rm peak}$ for $i=1,2,...$ and $n=\{3,3,5,5,10\}\times10^{-3}$ for each respective frequency. 15 and 23\,GHz, as well as 43 and 86\,GHz images have the same spatial scale.} 
 
    \label{fig:maps}
\end{figure}

Our new observations taken between 2017 and 2019 probe the innermost few milli-arcseconds of the 3C\,345 radio jet. Figure~\ref{fig:maps} shows exemplary total intensity CLEAN maps between 5 and 86\,GHz observed in May 2018, the month with the highest VLBI frequency coverage. 
The 5\,GHz observation took place on May 4 and was taken from the VLBA archive; the 15\,GHz MOJAVE image was observed on May 31, and the 23--86\,GHz data was observed on May 18.   

The total intensity morphology reveals a knotty jet on milli-arcsecond scales, which undergoes a northward bend some 5\,mas from the 15\,GHz VLBI core. We model the compact jet with circular Gaussian components using the \texttt{Difmap} task \texttt{MODELFIT} \citep{Pearson1994}. At 43\,GHz, the core, two stationary components S1 and S2, as well as seven moving jet components %
can be robustly identified in multiple consecutive epochs; the two stationary features are not distinguishable at 23\,GHz. Between 43\,GHz and 23\,GHz, components S1, Q1, and Q3--Q5 can be robustly cross-identified for more than three epochs. 
Component positions at 23, 43, and 86\,GHz of the new observations are listed in Table \ref{tab:compspos} for Q4 and are supplied as machine-readable table for all other components.

The central 0.5\,mas at 43\,GHz are well described by the (usually) easternmost component which we label the "core" in agreement with previous studies \citep{Schinzel2012,Weaver2022}, together with stationary components S1 and S2. For the first six epochs, S1 is the brightest feature instead of the 43\,GHz core; the latter becomes the brightest feature some time before May 18, 2018 upon ejection of a new component.

\begin{figure*}
	\centering
	\includegraphics[width=\textwidth]{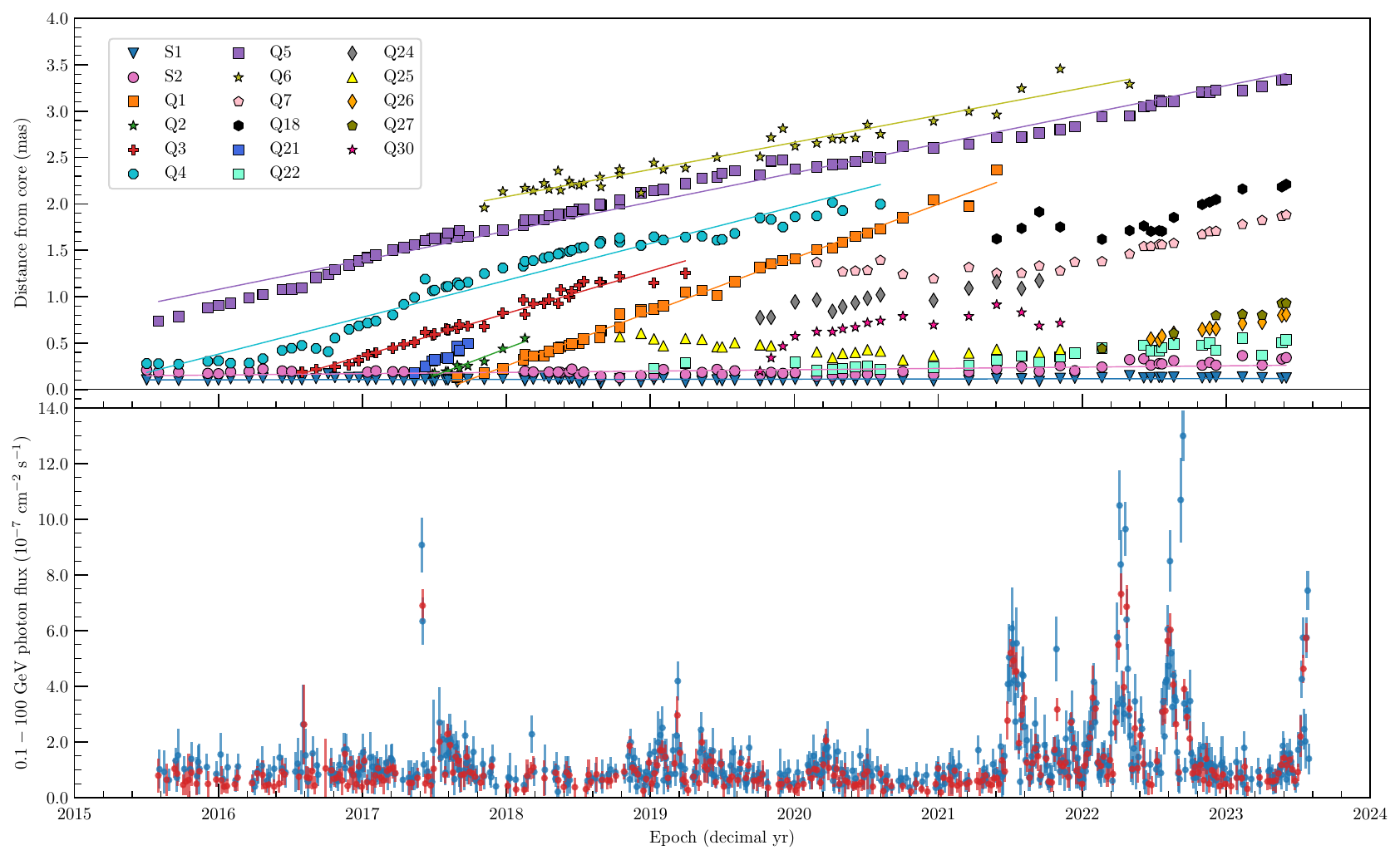}
	\caption{Core separation of robustly detected components in the 3C\,345 jet at 43\,GHz. The model fits from this study and \cite{Weaver2022} are shown and approximated by linear fits together to increase the robustness of the fits \edit{for the components identifiable across multiple frequencies} (top panel). Fit results are given in Table \ref{tab:kinematics}. FERMI-LAT gamma ray activity in weekly (red) and three-daily (blue) binning are shown in the bottom panel. The ejection of Q21, Q2, and Q23 has a gamma ray counterpart in mid-2017, while the later gamma activity cannot be identified with a clear-cut ejection event.}\label{fig:sep_evolution_Q}
\end{figure*}

The radial separation of Gaussian components at 43\,GHz with time is shown in Fig. \ref{fig:sep_evolution_Q} in an extended timerange between mid-2015 and mid-2023. Here, we improve the time sampling at 43\,GHz using model fits by \cite{Weaver2022}, shown together with the results from the observations described above.

\subsection{Jet kinematics}

Out of the eight persistent components mentioned in section \ref{sec:total_intensity}, 
we identify two stationary (S1 and S2) and seven superluminally moving components, 
similar to early VLBI studies of 3C\,345 \citep[e.\,g.][]{Zensus1995,Schinzel2012,Potzl2021} and similar sources. Newly ejected  components 
appear to pass through S2, appearing blended with it until they are separated far 
enough from the central engine. This leads to an ambiguity in the identification of 
Gaussian components and, in turn, to uncertainty whether or not new components are 
accelerating. For example, \cite{Schinzel2012} identified accelerating components 
during the 2009 gamma ray flare; however, \cite{Weaver2022} show those same components to move linearly instead. 

Since the time sampling at 43\,GHz exceeds that of the other frequencies by at least a factor of three, we carry out the kinematic analysis at 43\,GHz only. The apparent speeds of the moving components are summarized in Table \ref{tab:kinematics}. In good agreement with \cite{Weaver2022}, we find very high apparent velocities of $\sim$\,14--18\,c in the jet due to the very small viewing angle in 3C\,345. In fact, de-projecting the apparent speed $\beta_{\rm app}$ as \citep{Schinzel2012}
\begin{equation}
    \beta_{\rm depr}=\frac{\beta_{\rm app}}{\beta_{\rm app} \cos\theta + \sin\theta}, 
\end{equation}
and minimizing w.r.t. the viewing angle $\theta$, the de-projected speed is $\beta_{\rm depr}\gtrsim0.998$ for $\beta_{\rm app}>16$, and $\theta\lesssim 3.4$\degree.

\begin{figure}
	\centering
        \includegraphics[width=\columnwidth]{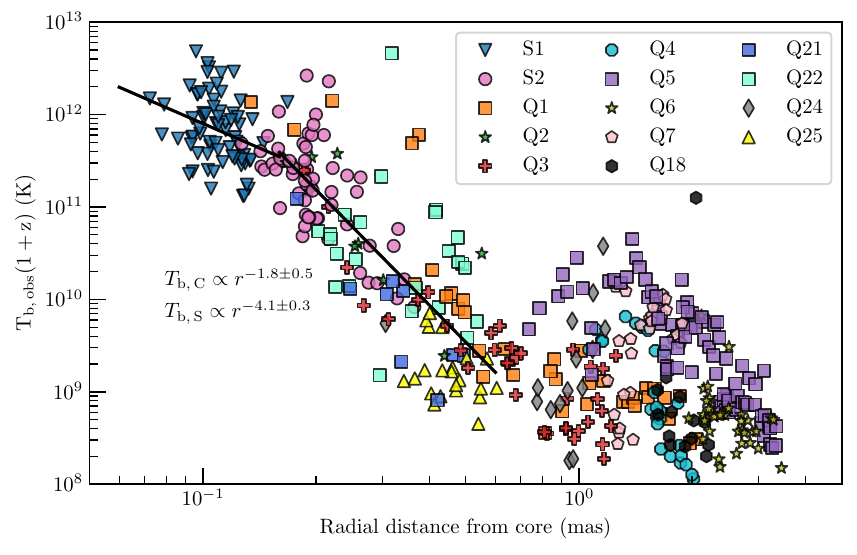}
	\caption{Brightness temperature evolution with apparent core separation of robustly detected components in the 3C\,345 jet at 43\,GHz. Results from the 43\,GHz observations \edit{from the 
 multiwavelength data set} discussed in this paper are shown together with results obtained by 
 \cite{Weaver2022} as well as model fits of the BEAM-ME project epochs from 2018 to 2023. The linear 
 fits (black lines) approximate regions below and above a "break" marking the transition from the 
 Compton- to the synchrotron-dominated losses at 0.16\,mas from the VLBI core, the average position 
 of S2 \citep{Weaver2022}. The slopes of the fits are annotated in the plot. We exclude unreasonably 
 high brightness temperatures above $5\times10^{12}$\,K.}
 
     \label{fig:tb_evolution_Q}
\end{figure}

\begin{figure*}[h]
	\centering 
	\includegraphics[width=\textwidth]{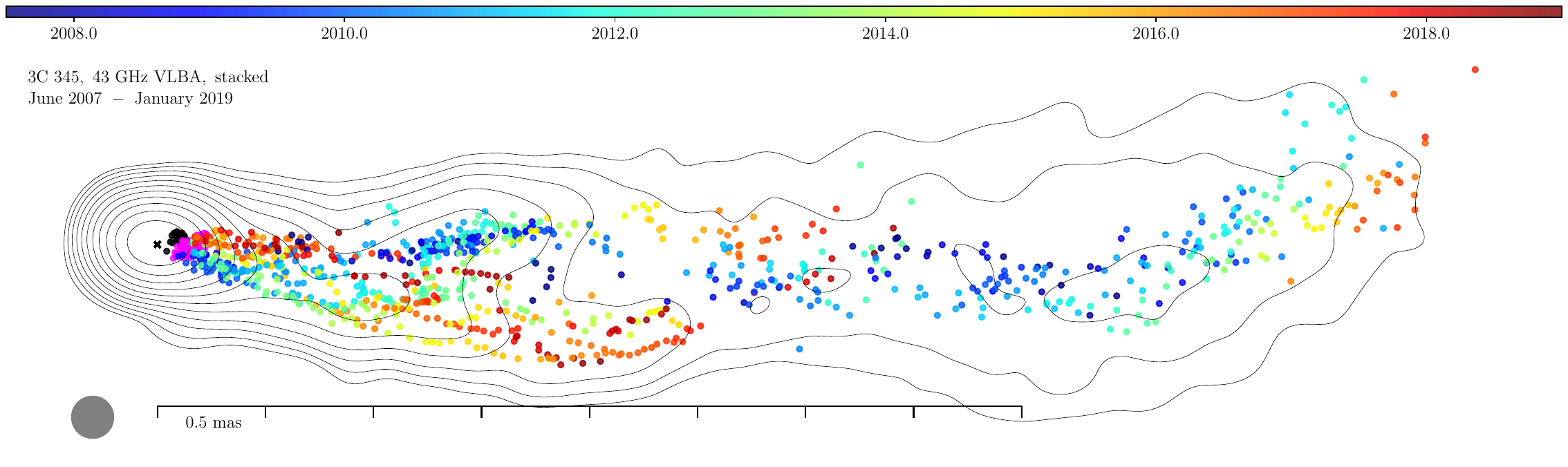}
    \caption{Long-term evolution of on-sky positions of model fit components. The component positions have been adjusted for the core to be located at the origin at all epochs. The background map is a stacked image of all epochs the components are taken from, restored with a 0.2\,mas circular beam, shown as grey circle in the bottom left corner. We restrict the time range to better illustrate the maxima of the jet slewing.}
    \label{fig:longterm}
\end{figure*}

The downstream components between 2017 and 2019 move along two distinct paths in the jet. Specifically, Q1 and Q2 follow the path of Q3 and Q4 past ejection, while Q5 and Q6 travel along a parallel track further south. We put this result into context with the long-term evolution by over-plotting the component positions obtained by \cite{Weaver2022} between June 2017 and January 2019 onto a stacked 43\,GHz image of the same time range, shown in Fig. \ref{fig:longterm}. The stationary features S1 and S2 are respectively shown in black and magenta to distinguish them from the moving components, shown color coded by the time of observation. The core is marked with a black cross.

Each newly ejected component starts its journey from an on-sky position coinciding with that of S2. Past this point, the 
stacked image jet opens up with a profile defined by the paths of moving components. While individual epochs tend to show a 
source morphology comparable to the 43\,GHz panel in Fig. \ref{fig:maps}, the stacked image reveals the underlying structure 
of the jet. It is hence evident that at any point in time, only some illuminated parts of the jet are visible. Therefore, the 
characteristic milli-arcsecond scale bends are not, in fact, "true" bends, but merely a consequence of components ejected in 
different directions traveling downstream in a given constellation.

Between mid-2007 and $\sim$\,2015, the jet completes about half of one slewing cycle, covering the entire "true" jet cross-
section with ejected components that continue to travel on curved paths. This behavior is broadly consistent with the 
$\sim$\,7.5\,yr activity cycle in 3C\,345 predicted by \cite{Lobanov1999}. The slewing is anchored at or before the second 
stationary feature S2 at a mean position of 0.16\,mas (1\,pc projected distance) from the core \citep{Weaver2022}. Assuming a 
viewing angle $\theta\sim 3$\degree, the region showing discernible component paths extends from $\sim$\,20\,pc to 
$\sim$\,380\,pc deprojected distance from the central engine. 

From the stacked image, it remains unclear whether the jet actually recollimates at $\sim$\,2.5--3\,mas from the 43\,GHz 
core, or if the apparent motion in a single funnel past this point is mimicked by sufficiently adiabatically expanded 
components. Indeed, model fit components past $\sim$\,2.5\,mas commonly reach FWHMs in excess of 1\,mas, filling out the jet 
cross-section at these scales, catching the faint extended emission. However, \cite{Schinzel2010} also identify components to be moving on multiple paths beyond the sharp northwards bend most visible at 15\,GHz a few milli-arcseconds from the core. 

\edit{The consistent movement of components over long periods of time supports a helical jet structure in 
3C\,345. Such geometries have been observed in a large number of AGN jets \citep[see, e.\,g.,][]{Reid1989,Owen1989,Zensus1995,Steffen1995,Zensus1997,Ostorero2004}. They may be caused by instabilities 
in the flow \citep{Fuentes2023,Nikonov2023}, jet precession \citep{Lu1990,Britzen2023,vonFellenberg2023}, 
or even be the signature of a binary black hole in the central engine \citep{Villata1999,Lobanov2005}. Taken by themselves, the 
data presented in this paper clearly indicate a curved morphology that changes somewhat 
periodically in time. Together with the magnetic field structure (see Sect. \ref{sec:polarization}) and the 
results obtained in previous studies, this provides strong evidence for a helical jet flow.}

\subsection{Brightness temperatures}

We estimate the brightness temperature $T_{\rm b}$ as
\begin{equation}\label{eq:Tb}
	T_{\rm b}=\frac{c^2}{2k_B\nu^2}\frac{4 \ln 2}{\pi r^2} (1+z) \ S_\nu,
\end{equation}
where $S_\nu$ and $r$ the component flux density in Jy and radius in mas, respectively. Since this 
prescription differs from that used by \cite{Weaver2022} by a multiplicative factor, we re-evaluate 
their $T_{\rm b}$ results for our analysis. Figure \ref{fig:tb_evolution_Q} shows the brightness 
temperature evolution of the robust components at 43\,GHz with increasing on-sky distance from the 
VLBI core. 

We approximate the data by a broken power-law below and above the average position of S2 at 
0.16\,mas from the core \citep{Weaver2022}, yielding slopes of $-1.9\pm0.5$ and $-4.4\pm0.3$ 
respectively. Taking into account the large scatter in the brightness temperature, partly caused by 
small component sizes in some epochs, we confirm the result obtained by \cite{Schinzel2012}: up 
until the second stationary feature, the jet is dominated by Compton losses, and synchrotron losses 
take over beyond that point. For a straight jet, the slope would soften lightly towards the end of 
the synchrotron stage to make way for adiabatic expansion to dominate radiative losses. However, in 
3C\,345 the data points start to scatter heavily around 1\,mas from the core, where the helical 
paths in the jet cross for the first time after the initial jet opening (see Fig. 
\ref{fig:longterm}).

\subsection{Origin of the gamma-ray emission}

\cite{Schinzel2012} have shown that relativistic outflow accounts for at least parts of the observed 
gamma-ray emission. Specifically, they pinpointed the emission site to the component closest to the 
43\,GHz VLBI core, which we confirm to be Compton-loss dominated. Since the flare of 2009 was 
coinciding with the ejection of multiple new components into the jet, one may suspect a shock-shock 
interaction scenario to be the underlying cause of the gamma-ray emission. This would imply a spike 
in the gamma rays once a newly formed component passes through a stationary feature, which are 
commonly identified to be standing shocks. However, Fig. \ref{fig:sep_evolution_Q} shows that the 
connection of increased gamma ray activity and radio jet evolution is not as clear-cut: For example, 
the gamma spike in mid-2017 is small, but appears to correspond to three components crossing a 
standing feature close in time. In contrast, the ejection event in mid-2016 appears to not have any 
gamma-ray counterpart, while the much higher gamma activity between 2021 and 2023 does not seem to 
correspond to the ejection of any individual component. 

\begin{figure}
	\centering
        \includegraphics[width=\columnwidth]{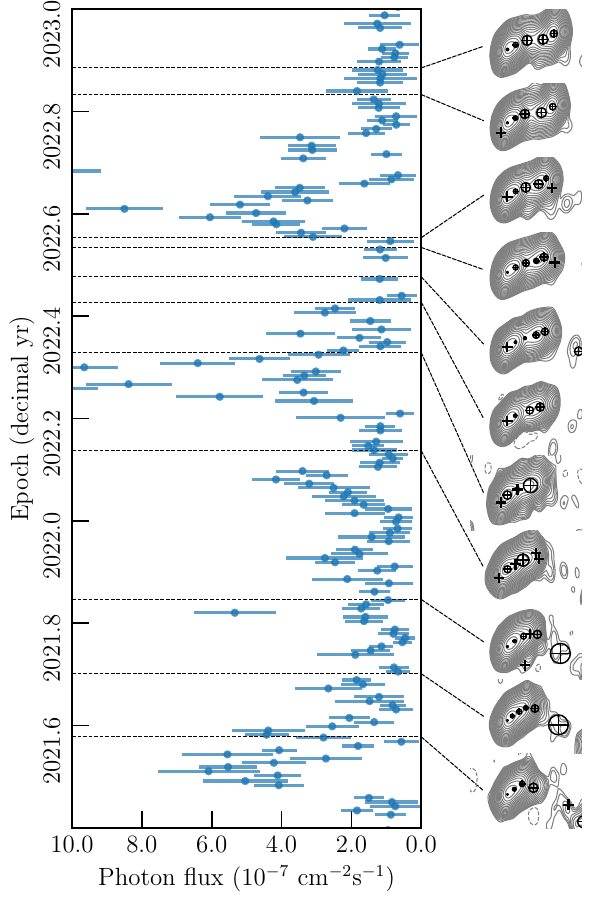}
	\caption{Fermi-LAT gamma ray light curve (left panel) and 43\,GHz VLBI total intensity maps taken from the BUBG BEAM-ME project (right panels). Circular Gaussian model fit components are overplotted as black circles or black crosses depending on their size. The VLBI maps are aligned to the modelfit core. }
     \label{fig:gammaVLBI}
\end{figure}

Between November 2021 and November 2022, a feature comparable in size to the VLBI core slowly 
separates from it, shown on the right side of Fig. \ref{fig:gammaVLBI}. The accuracy of the model 
fit during that time is heavily decreased. It is likely that during this event, different parts of 
the jet violently interact with each other, producing the extreme gamma ray activity. The feature is 
also highly polarized and governed by a transverse magnetic field, indicating that it consists of 
moving shocks like the other traveling components. It is unclear which interaction between the 
components is causing the extreme gamma ray activity, shown on the left side of Fig. \ref{fig:gammaVLBI}.

\subsection{Core shift}

For a classical Blandford-Königl AGN jet, the position of the optically thick 
VLBI core with respect to the optically thin jet depends on the observing 
frequency; hence, the core appears displaced when overlaying images of the same 
jet at two different frequencies. The magnitude of this core shift is a measure 
of the system's deviation from equipartition, i.\,e., whether a larger portion of 
the total energy is stored in the magnetic field, or in the particles. The core 
position depends on the frequency as $r_{\rm core} \propto \nu_{\rm obs}^{-1/k_r}$ \citep{Konigl1981}, where for a system in equipartition between 
magnetic and particle energy, $k_r=1$. The magnetic field and the particles 
respectively dominate for $k_r>1$ and $k_r<1$. 

In order to determine the core shift, we first mask the optically thick core 
region in two maps with the same pixel size, restored with the same beam. Then, 
we Fourier-align the images via 2D cross-correlation. 

In May 2018, the frequency coverage is sufficient to perform a power-law fit to 
the VLBI core positions from 5 to 86 GHz. \edit{Defining an offset 
$\Omega_{r\nu}$, the true core position reads \citep{Lobanov1998}:
}
\begin{equation} \label{eq:cshift_plaw}
    r_{\rm core}(\nu)=\Omega_{r\nu} \left[ \nu^{1/k_r} \sin\theta  \right]^{-1}.
\end{equation}
We fit Eq. \ref{eq:cshift_plaw} to the obtained core shifts in May 2018 
solving for the pre-factor $\Phi=\Omega_{r\nu}/\sin\theta$ and exponent 
$k_r$ as $r_{\rm c}\sim \Phi \times \nu^{-1/k_r}$.

\begin{figure}
    \centering
	\includegraphics[width=\columnwidth]{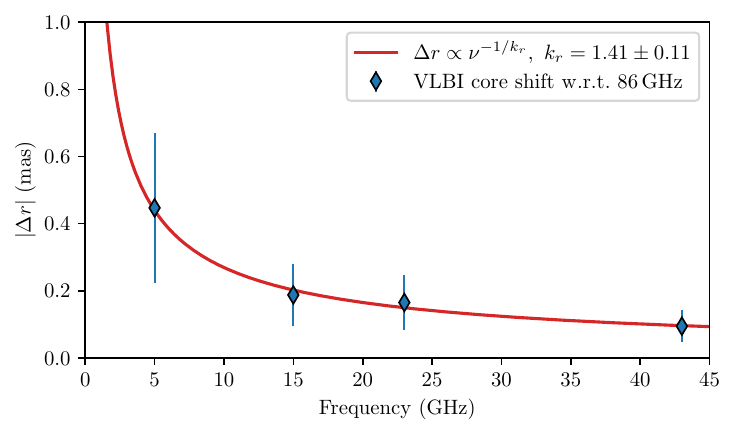}
	\caption{Radial core shift measurement in the May 2018 epoch with respect to the core at 86\,GHz, based on the total intensity maps shown in Fig. \ref{fig:maps}.} 
    \label{fig:coreshift}
\end{figure}

Figure \ref{fig:coreshift} shows the \edit{radial core shift} at a given 
frequency with respect to the 86\,GHz map. The 5/15\,GHz pair is set to 
a 0.02\,mas pixel size and is convolved with a 1.6\,mas beam, whereas 
the maps in all other frequency pairs are set to 0.01\,mas pixels and a 
0.4\,mas beam. The shifts are then obtained using the masking and 
alignment procedure described above. The power law fit yields 
$k_r\simeq1.4$, implying a magnetically dominated jet.

During the period from 2017-2019, we study the evolution of the core shift 
between 23 and 43\,GHz. In the third epoch of our multi-frequency observations, 
the core shift drops from the mean value $|\Delta r|=(0.055\pm0.019)$\,mas down 
to 0.028\,mas, corresponding to the ejection of three radio features and a spike 
in the gamma ray emission during that time frame (see Fig. 
\ref{fig:sep_evolution_Q}). The scatter around the mean in the other, quiescent 
epochs, is likely caused by uncertainties in the alignment.

\subsection{Spectral index}\label{sec:spectralindex}

We obtain spectral index maps from the total intensity maps for each epoch between 23 and 43\,GHz ($\alpha_{23-43}$), adopting the definition $S\propto\nu^\alpha$ and accounting for the core shift. 

Since the VLBA has been known to have issues in total flux calibration in recent years\footnote{\href{https://www.cv.nrao.edu/MOJAVE/}{https://www.cv.nrao.edu/MOJAVE/}}, we re-scale the total intensity maps to fit the 
single-dish spectral index at corresponding epochs using the OVRO 15\,GHz and Metsähovi 
37\,GHz light curves. The 23\,GHz maps are upscaled $\sim$\,25\,\%, while the 43\,GHz 
maps require a multiplicative factor of 1.40 on the visibility amplitudes. Satisfactory 
spectral index maps between 43\,GHz and 86\,GHz can not be obtained, since the correct 
scaling factor for the 86\,GHz data cannot be reliably determined (scaling to ALMA band 
3 flux densities would yield scaling of a factors of four or more, likely influenced by 
extended emission that is resolved out in VLBI). Additionally, only very little jet 
structure is recovered at 86\,GHz, hindering image alignment via 2D cross-correlation.

Between 23\,GHz and 43\,GHz, the core shows a flat or lightly inverted spectrum (${\alpha_{23-43}\sim0.25}$). 
In the jet, generally ${\alpha_{23-43}\lesssim0}$; for some epochs, the jet edges partly show inverted spectra, 
which may be identified as artifacts due to the post-imaging flux scaling. 

\begin{figure}
    \centering
	\includegraphics[width=\columnwidth]{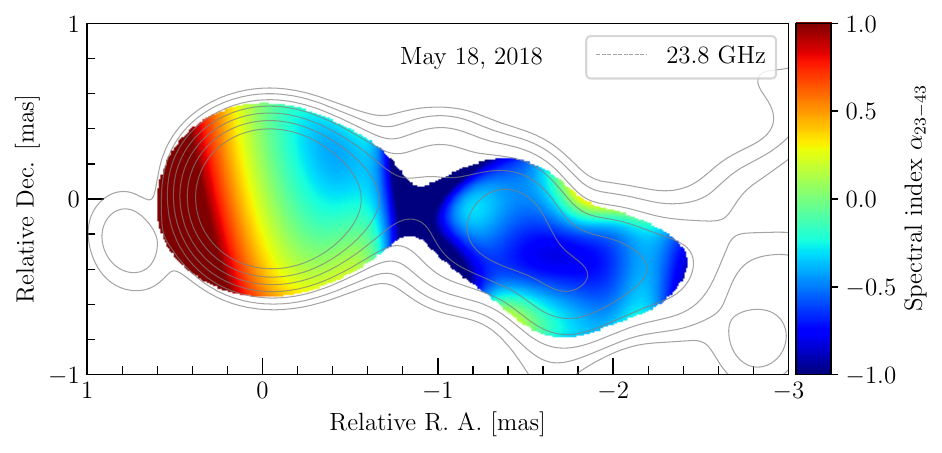}
	\caption{Spectral index map between 23\,GHz (in contours) and 43\,GHz for a selected epoch. Both maps are restored with a 0.4\,mas FWHM circular beam.} 
    \label{fig:spix}
\end{figure}

\edit{Figure \ref{fig:spix} shows an exemplary spectral index map between 23\,GHz and 43\,GHz on May 18, 2018.} 
At $\sim$\,1\,mas from the core, the spectral index maps show a consistent, 
stationary minimum. The 23\,GHz jet appears continuous, but is very faint at 
43\,GHz and even shows a gap in some epochs (see Fig. \ref{fig:maps}). This gap 
apparently coincides with a break in the brightness temperature evolution along 
the jet at 43\,GHz as described above \edit{(see Fig. \ref{fig:tb_evolution_Q})}. The long-term evolution (see Fig. \ref{fig:longterm}) 
reveals that the helical paths of ejected components appear 
to cross at this distance from the core, indicating that the flux density in 
this region \edit{may be} decreased at 43\,GHz due to a change in viewing angle. 

\edit{This also 
supports a picture consistent with Sect. \ref{sec:polarization}: at observing frequencies ${\lesssim15}$\,GHz, 
for the most part, we are probably looking to the "outer envelope" of the jet. At frequencies $\gtrsim43$\,GHz, with a 
transition somewhere in between, the radiation permeates this envelope and reveals the structure of the inner 
jet. While in the case of a steady jet this may reveal a form of "spine", possibly in the form of an edge 
brightening towards higher frequencies, for a jet as knotty as the one in 3C\,345 it instead allows us to 
follow the paths of moving features within the jet. }

The region around the bend between 1 and 2\,mas from the core shows a moderate 
spectral index of $\sim$\,--0.5, as is usually expected in AGN jets. Toward the 
jet edges, the index generally slightly increases. This may be indicative of a 
shear layer which is visible at 23\,GHz and is resolved out at higher 
frequencies, revealing the traveling shocks inside the jet.

\subsection{Polarized structure}\label{sec:polarization}

\begin{figure}[h]
    \centering
    \includegraphics[width=\columnwidth]{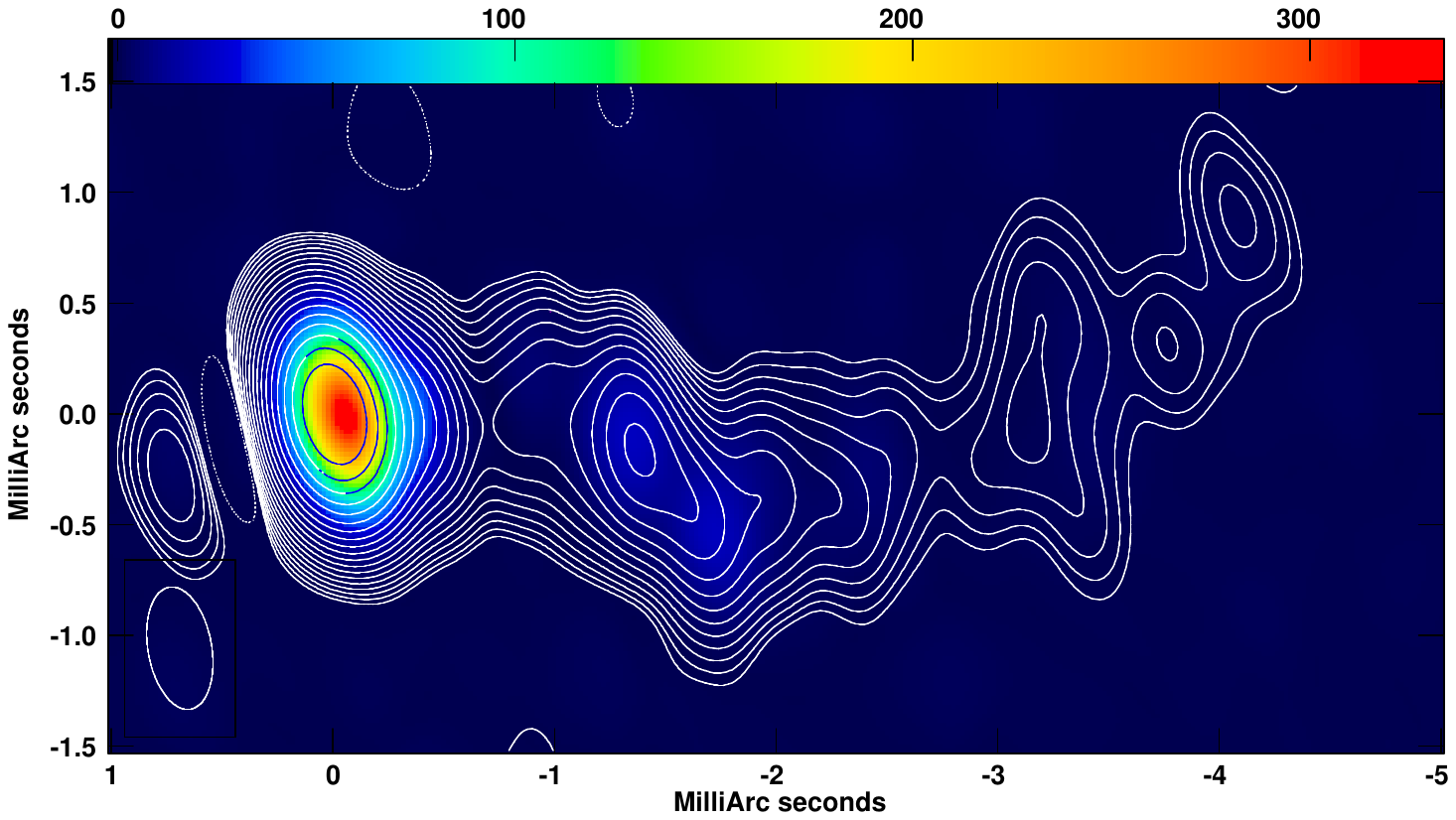}
    \includegraphics[width=\columnwidth]{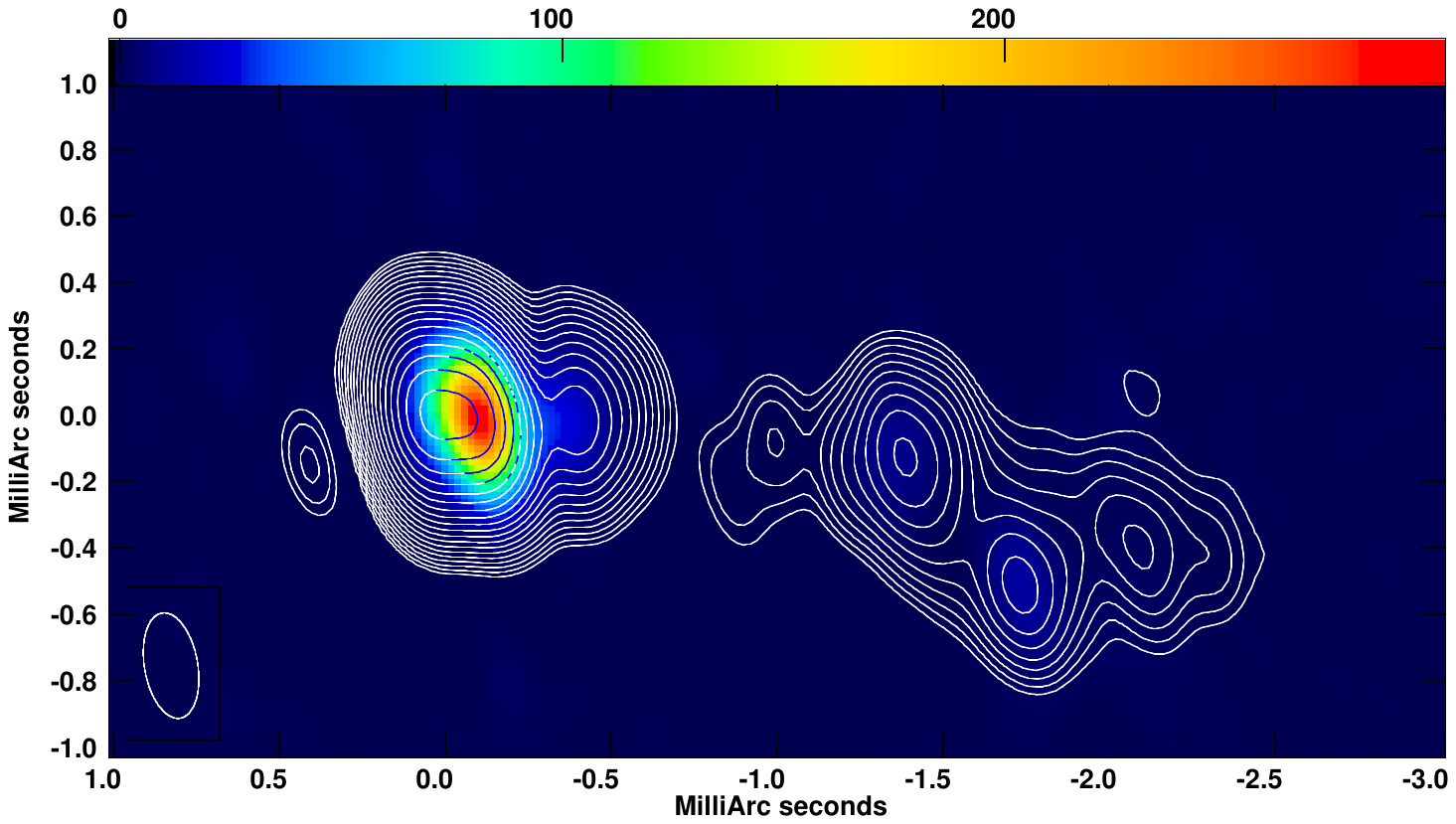}
    \includegraphics[width=\columnwidth]{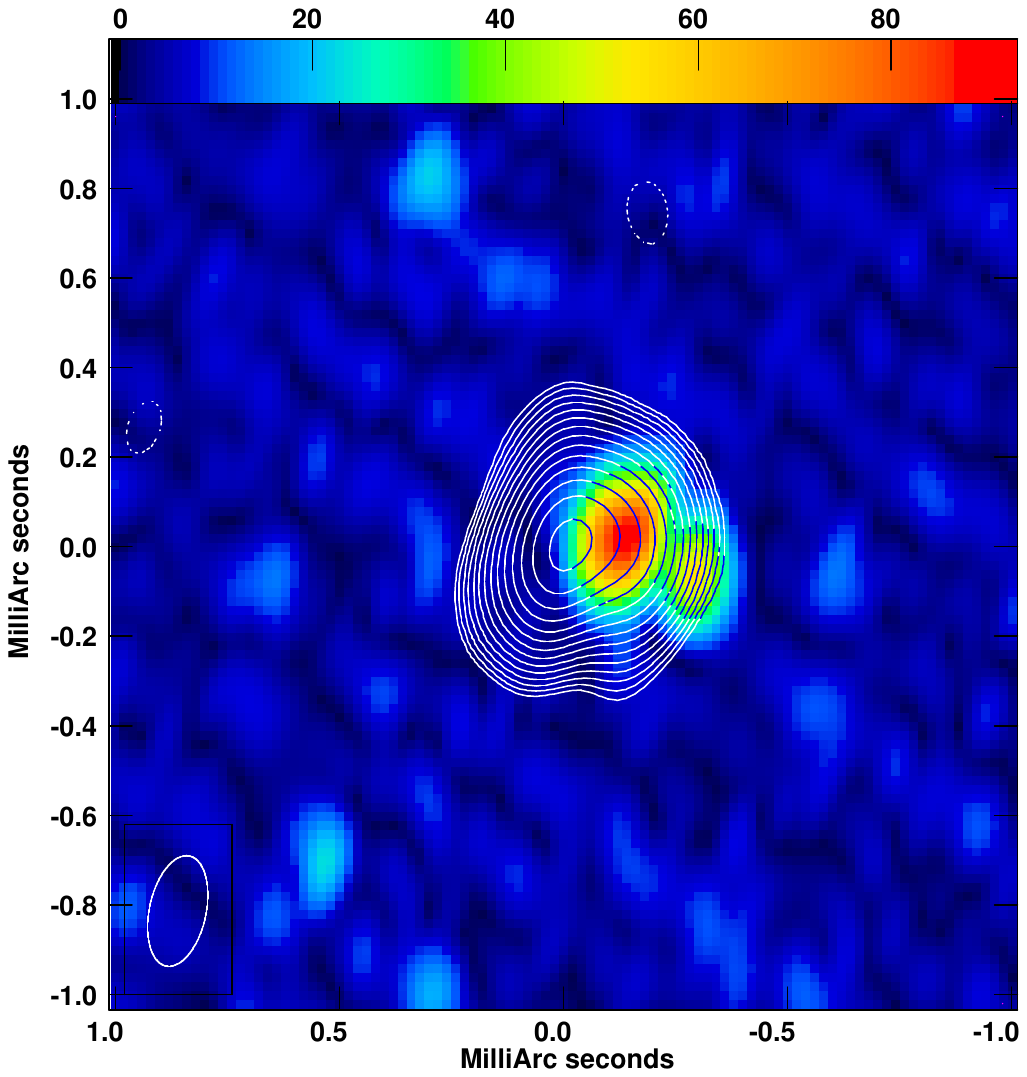}
    \caption{Total (contours) and polarized intensity (color map, in 
    mJy/beam) on May 18, 2018. The base contour levels are 0.005 for 
    23 and 43\,GHz (top and middle panels), and 0.02 for 86\,GHz 
    (bottom panel). The polarized maximum is consistently displaced 
    from the total intensity maximum and rather coincides with the 
    positions of the stationary features. In the lower two 
    frequencies, the left and middle of the moving shocks around 1 
    to 1.5\,mas from the core are weakly polarized.}
    \label{fig:3Cpol}
\end{figure}

At 23, 43, and 86\,GHz, the polarized maximum is located near that of the total 
intensity map, with a small downstream displacement towards the stationary 
components S1 and S2 \edit{(see Fig. \ref{fig:3Cpol})}. For a given epoch, the 
total polarized intensity decreases monotonously with decreasing frequency. While 
at 23\,GHz the polarized maximum reaches 400\,mJy/beam, at 86\,GHz it reaches 
about one tenth of that. Out of the traveling components, Q4 and Q5 show low, but 
consistently recovered polarization of $\lesssim 50 $\,mJy/beam at 23\,GHz and 
43\,GHz. This 43\,GHz structure even carries over to stacked images of the source 
\citep{MacDonald2017}.   

In the three MOJAVE 15\,GHz epochs between 2017 and 2019, 3C\,345 shows a 
fractional linear polarization $m\lesssim 0.1$ up until $\sim$\,10\,mas from the 
core. In some downstream regions starting at $\sim$\,5\,mas, $m\lesssim0.3$ is 
reached at the jet edges. 
The EVPAs start off parallel to the core, but abruptly 
change their orientation from the jet base outward. 
Comparing to the 43\,GHz 
morphology, the core retains its transverse magnetic field across frequencies. 
The traveling downstream shocks, however, no longer show a transverse field at 
15\,GHz as they do at 43\,GHz and as expected from shock compression 
\citep{Laing1980,Hughes1985}. This indicates the presence of a shear layer 
between the jet and the ambient medium seen at 15\,GHz that is resolved out at 
higher frequencies.

\section{Summary, conclusions and outlook}

In this work we have analyzed ten epochs of VLBA observations at 22, 43, and 86\,GHz of the famous quasar 3C\,345, complemented by 87 epochs at 43\,GHz taken from the VLBA-BU-BLAZAR and BEAM-ME programs. The bent compact jet, characteristic for 3C\,345, consists of two stationary and six superluminally moving components that can be identified across epochs and frequencies between 2017 and 2019. 

\begin{enumerate}
    \item[$\bullet$] The innermost $\sim$\,0.6\,mas of the 43\,GHz jet are dominated by Compton and synchrotron losses; the transition is marked by a stationary feature some 0.16\,mas from the core. 

    \item[$\bullet$] The long-term 43\,GHz evolution reveals components moving on helical paths in the jet. Points at which the paths apparently cross appear faint, causing a steep spectral index and a flux density gap in the total intensity images. 

    \item[$\bullet$] The gamma-ray emission partly coincides with the ejection of new jet components, but can also be related to less specific but extreme changes in the jet morphology, as can be seen during the period of high gamma-ray activity from 2021 to 2023.

    \item[$\bullet$] In the quiescent May 2018 epochs, the core shift indicates a magnetically dominated jet.

    \item[$\bullet$] \edit{Long-term component tracking shows that the characteristic bent jet represents the part of the jet illuminated by moving components at a given time of observation, rather than a "true" bend in the flow caused by external factors. The stacked image reveals the full jet cross-section.}
    

    \item[$\bullet$] The polarization structure in conjunction with the component movement at 43\,GHz indicate that up to $\sim$\,23\,GHz, mostly an outer sheath of the jet is observed, which is resolved out moving to higher frequencies. Specifically, at 15\,GHz the EVPA pattern suggests interaction with the ambient medium, \edit{whereas the 43\,GHz polarization structure reveals an EVPA (and magnetic field) structure consistent with the moving shocks.} 

\end{enumerate}

\remove{The results from this study will serve as inputs for a 
magneto-hydrodynamic (MHD) simulation follow-up work. Specifically, 
we aim to investigate the connection of shock-shock interactions in 
the jet to the gamma-ray emission and the movement of shocks along 
curved paths.}

\hfill{\fontfamily{calligra}\selectfont  The End...?}

\begin{acknowledgements}
We thank Luca Ricci, Petra Benke, Manel Perucho and Christian Fromm for their support in discussions on the results, and Phil Hughes in his role as the journal referee.

JR received financial support for this research from the International Max Planck Research School (IMPRS) for Astronomy and Astrophysics at the Universities of Bonn and Cologne. 

This research is supported by the European Research Council advanced grant “M2FINDERS - Mapping Magnetic Fields with INterferometry Down to Event hoRizon Scales” (Grant No. 101018682). 

The VLBA is an instrument of the National Radio Astronomy Observatory. The National Radio Astronomy Observatory is a facility of the National Science Foundation operated by Associated Universities, Inc. 

This study makes use of VLBA data from the VLBA-BU Blazar Monitoring Program (BEAM-ME and VLBA-BU-BLAZAR\footnote{\href{http://www.bu.edu/blazars/BEAM-ME.html}{http://www.bu.edu/blazars/BEAM-ME.html}}), funded by NASA through the Fermi Guest Investigator Program. 



\\

{\it Software.} {\tt AIPS}\footnote{\href{http://www.aips.nrao.edu/index.shtml}{http://www.aips.nrao.edu/index.shtml}} \citep{Greisen2003}, {\tt Difmap}\footnote{\href{https://www.cv.nrao.edu/adass/adassVI/shepherdm.html}{https://www.cv.nrao.edu/adass/adassVI/shepherdm.html}} \citep{Shepherd1997}, {\tt ehtim}\footnote{\href{https://achael.github.io/eht-imaging/}{https://achael.github.io/eht-imaging/}} \citep{Chael2018a}
\end{acknowledgements}

\bibliographystyle{aa}

\bibliography{3C345}

\appendix 

\section{Supplementary material}

Table \ref{tab:compspos} shows the obtained fluxes and positions  for a representative jet component. Table \ref{tab:obssummary} provides an overview of the ten multi-frequency epochs observed between 2017 and 2019. Total fluxes at 23 and 43\,GHz are scaled to the single-dish spectral index between 15 and 37\,GHz.  

\begin{table}[h]
	\caption{Component positions}
	\label{tab:compspos}
	\centering
        \begin{tabular}{cccccc}
			\hline\hline
Epoch & ID & Flux & X & Y & FWHM \\
 &  & Jy & mas & mas & mas \\\hline
\multicolumn{6}{c}{23\,GHz} \\ 
2017.486 & 4 & 0.188 & --1.031 & --0.069 & 0.314  \\ 
2017.580 & 4 & 0.208 & --1.118 & --0.118 & 0.304  \\ 
2017.659 & 4 & 0.211 & --1.137 & --0.127 & 0.275  \\ 
2017.733 & 4 & 0.181 & --1.136 & --0.099 & 0.240  \\ 
2018.116 & 4 & 0.308 & --1.250 & --0.134 & 0.272  \\ 
2018.262 & 4 & 0.355 & --1.392 & --0.139 & 0.239  \\ 
2018.377 & 4 & 0.233 & --1.389 & --0.136 & 0.063  \\ 
2018.437 & 4 & 0.335 & --1.443 & --0.128 & 0.234  \\ 
2018.505 & 4 & 0.359 & --1.481 & --0.150 & 0.272  \\ 
2018.656 & 4 & 0.325 & --1.552 & --0.156 & 0.282  \\ \hline
\multicolumn{6}{c}{43\,GHz} \\ 
2017.486 & 4 & 0.076 & --1.062 & --0.067 & 0.166  \\ 
2017.580 & 4 & 0.087 & --1.106 & --0.072 & 0.139  \\ 
2017.659 & 4 & 0.090 & --1.069 & --0.060 & 0.162  \\ 
2017.733 & 4 & 0.093 & --1.153 & --0.078 & 0.166  \\ 
2018.116 & 4 & 0.231 & --1.325 & --0.121 & 0.193  \\ 
2018.262 & 4 & 0.234 & --1.410 & --0.134 & 0.210  \\ 
2018.377 & 4 & 0.237 & --1.464 & --0.145 & 0.223  \\ 
2018.437 & 4 & 0.230 & --1.482 & --0.142 & 0.215  \\ 
2018.505 & 4 & 0.217 & --1.520 & --0.133 & 0.221  \\ 
2018.656 & 4 & 0.204 & --1.590 & --0.147 & 0.276  \\\hline
\multicolumn{6}{c}{86\,GHz} \\ 
2017.486 & 4 & -- & -- & -- & --  \\ 
2017.580 & 4 & 0.019 & --1.317 & --0.052 & 0.047  \\ 
2017.659 & 4 & -- & -- & -- & --  \\ 
2017.733 & 4 & -- & -- & -- & --  \\ 
2018.116 & 4 & 0.025 & --1.282 & --0.173 & 0.293  \\ 
2018.262 & 4 & -- & -- & -- & --  \\ 
2018.377 & 4 & 0.028 & --1.466 & --0.175 & 0.098  \\ 
2018.437 & 4 & 0.013 & --1.502 & --0.215 & 0.010  \\ 
2018.505 & 4 & 0.010 & --1.339 & --0.194 & 0.374  \\ 
2018.656 & 4 & 0.066 & --1.422 & --0.378 & 0.622  \\
\hline 
   		\end{tabular}

	\tablefoot{Excerpt from Gaussian model fit component positions for the multifrequency VLBA dataset, shown for component Q4 for a selected time frame. The full table including all components is supplied in machine-readable format.}
\end{table}

\begin{table*}
	\caption{List of multi-frequency VLBA observations. }
	\label{tab:obssummary}
	\centering
		\begin{tabular}{cccc lll c}
			\hline\hline

	Date & \multicolumn{3}{c}{$S_{\rm tot}$ (Jy)} & \multicolumn{3}{c}{Beam (bpa), ($\upmu$as)$\times$($\upmu$as) ($^\circ$)} & Missing \\ 
        \cmidrule(l){2-4}  \cmidrule(l){5-7}
	YYYY-MM-DD& 23\,GHz & 43\,GHz & 86\,GHz & 23\,GHz & 43\,GHz & 86\,GHz & Antennas \\ \hline
   
    2017-06-27 & 5.60  & 6.02  & 0.75  &  $532\times272\ (-3)$  &  $285\times151\ (1)$   & $265\times88 \ (-18)$   &     \\ 
	2017-07-31 & 5.97  & 6.88  & 1.43  &  $498\times251\ (-5)$  &  $276\times147\ (-4)$  & $283\times94 \ (-20)$   & FD  \\ 
	2017-08-29 & 6.17  & 6.97  & 1.09  &  $510\times286\ (-1)$  &  $288\times161\ (3)$   & $235\times104 \ (-24)$  &     \\ 
	2017-09-25 & 5.99  & 6.20  & 0.90  &  $750\times312\ (-16)$ &  $405\times185\ (-13)$ & $250\times125 \ (-15)$  & SC  \\ 
	2018-02-12 & 6.66  & 6.73  & 1.09  &  $768\times338\ (-13)$ &  $434\times199\ (-12)$ & $297\times230 \ (3)$    & HN, SC  \\ 
	2018-04-06 & 6.48  & 6.18  & 0.32  &  $555\times265\ (-8)$  &  $300\times154\ (1)$   & $310\times224 \ (-42)$  &     \\ 
	2018-05-18 & 6.20  & 6.12  & 1.46  &  $583\times275\ (0)$   &  $307\times156\ (4)$   & $270\times110 \ (-14)$  & HN  \\ 
	2018-06-09 & 6.11  & 6.24  & 1.02  &  $571\times309\ (9)$   &  $315\times159\ (10)$  & $349\times210 \ (13)$   &     \\ 
	2018-07-04 & 6.01  & 6.01  & 0.86  &  $515\times256\ (-2)$  &  $277\times149\ (2)$   & $525\times493 \ (38)$   & PT  \\ 
	2018-08-28 & 5.70  & 5.88  & 0.79  &  $532\times286\ (-5)$  &  $277\times151\ (4)$   & $294\times154 \ (7)$    &     \\
	
            \hline 
   		\end{tabular}

	\tablefoot{
   $S_{\rm tot}$: total flux density recovered in VLBA image clean map; Beam (bpa): beam size, major axis vs minor axis with position angle of ellipse in parentheses. In addition to the listed missing antennas, HN and SC lack an 86\,GHz receiver. }
\end{table*}

\begin{table*}
	\caption{Derived physical parameters for the components cross-identified between 23 and 43\,GHz components.}
	\label{tab:kinematics}
	\centering
			\begin{tabular}{l llllll}
				\hline\hline
				
				Label  & Q\,1 (B15) & Q\,2 (B14) &  Q\,3 (B12) & Q\,4 (B11) & Q\,5 (B10) & Q\,6 (C3) \\ \hline

    \# & 40 & 10 & 36 & 63 & 85 & 37  \\ 
$\langle\mu\rangle$ [mas/yr] & 0.58$\pm$0.01 & 0.59$\pm$0.05 & 0.46$\pm$0.01 & 0.40$\pm$0.01 & 0.31$\pm$0.00 & 0.29$\pm$0.01  \\ 
$\langle\beta_{\rm app}\rangle$ [c] &  20.03$\pm$0.29 & 20.21$\pm$1.70 & 15.80$\pm$0.50 & 13.70$\pm$0.50 & 10.83$\pm$0.16 & 10.10$\pm$0.50 \\

\hline

                $\langle\beta_{\rm app}\rangle^{\star}$ [c] &  $20.0\pm1.7$ &$20.2\pm2.6$ & $15.8\pm0.5$ & $15.8\pm0.8$& $14.9\pm0.4 $& $13.6\pm1.9$\\
				$\Gamma_{\rm var}^{\star}$ &$20.2\pm1.9$& $20.3\pm2.8$& $16.7\pm0.8$& $58.0\pm23.0$ & $24.0\pm3.0$ & $53.0\pm19.0$ \\
               $\delta_{\rm var}^{\star}$ &$18.0\pm3.0$ &$18.7\pm2.5 $& $22.0\pm 2.8$& $2.2\pm0.9$  & $5.2\pm0.8$ & $103.9\pm37.8$ \\ 
				$\theta_{\rm var}^{\star}$ &$3.2\pm0.5$ & $3.1\pm0.4$&$2.5\pm0.5$ & $7.2\pm0.4$ & $6.87\pm 0.26$ &  $0.2 \pm 0.1$\\
				\hline
				
			\end{tabular}

		\tablefoot{$^\star$Apparent speed, variability Doppler factor, variability Lorentz factor and viewing angle taken from \cite{Weaver2022}. \#: No. of epochs used for linear fit. Labels in brackets follow the naming convention by \cite{Weaver2022}, who sort the components after the ejection epoch instead of the increasing distance from the core in a particular snapshot.}
\end{table*}


\end{document}